# Broad Band Photon Harvesting Biomolecules for Photovoltaics

P. Meredith*

B. J. Powell, J. Riesz, R. Vogel, D. Blake

*The University of Queensland School of Physical Sciences, Soft Condensed Matter Physics Group & Centre for Biophotonics and Laser Science*

I. Kartini

*The University of Queensland School of Engineering & ARC Centre for Functional Nanomaterials*

G. Will, S. Subianto

*Queensland University of Technology School of Physical Sciences, Centre for Instrumental and Developmental Chemistry*

\* Corresponding author address:
*University of Queensland Department of Physics, St. Lucia Campus, Brisbane, QLD 4072 Australia*
*Email:*Meredith@physics.uq.edu.au

**Abstract**

In this chapter, we discuss the key principles of artificial photosynthesis for photovoltaic energy conversion. We demonstrate these principles by examining the operation of the so-called "dye sensitized solar cell" (DSSC) – a photoelectrochemical device which simulates the charge separation process across a nano-structured membrane that is characteristic of natural systems. These type of devices have great potential to challenge silicon semiconductor technology in the low cost, medium efficiency segment of the PV market. Ruthenium charge transfer complexes are currently used as the photon harvesting components in DSSCs. They produce a relatively broad band UV and visible response, but have long term stability problems and are expensive to manufacture. In this chapter, we suggest that a class of biological macromolecules called the melanins may be suitable replacements for the ruthenium complexes. They have strong, broad band absorption, are chemically and photochemically very stable, can be cheaply and easily synthesized, and are also bio-available and bio-compatible. We demonstrate a melanin-based regenerative solar cell, and discuss the key properties that are necessary for an effective broad band photon harvesting system.

# 1. Introduction

In the last chapter, we saw how and why silicon-based solid-state photovoltaics remain dominant in the solar cell market. We also saw that even the most advanced "*third generation*" silicon devices are based upon the conventional *pn* semiconductor junction. As such, the photon absorption and charge separation events occur at the same point in space within the depletion region. The electric field that drives the spatial separation of the electron-hole pair is essentially derived from the differences in electron affinities between the *n* and *p* type materials. This correspondence between the photon absorption and charge separation events is a key point of differentiation between the photovoltaic effect in a semiconductor junction, and the photon induced generation of a chemical potential in natural systems, i.e. photosynthesis. In the latter, and this is a very simplistic but highly relevant interpretation in the context of artificial photosynthetic systems, the point in space at which the primary light absorption event occurs can be very different from the point at which the "effect" is ultimately felt through the generation of spatially separated charge carriers. Whether or not photon absorption leads to the production of electricity or the initiation of some useful chemical reaction, removal of the "*spatial correspondence*" constraint has a profound effect upon device design – namely, we are not forced into providing a local electric field at the point of initial photo-excitation.

If we take as a classic example the PSI and PSII photosystems of plants, we see that dozens of chlorophyll molecules within the Light Harvesting Complex are capable of the primary absorption event. This energy is subsequently transferred with near unity quantum efficiency to a reaction-centre chlorophyll in the photosystem complex, where it can be released as a pair of spatially separated charge carriers. Clearly, the separation event requires some local electric field (as in the case of the *pn* junction), but that field need not exist at the absorption site. The key advantage of this strategy is clear: the tasks of photon absorption and charge separation can be decoupled, and therefore different molecules can be engineered and optimized for each function. It also means that a very large number of molecules can be used for photon harvesting (presenting a substantial total surface area for light absorption), whilst maintaining a relatively small number of specialist reaction-centres to facilitate charge separation.

In this chapter, we will explore this decoupling concept from the point of view of bio-mimetic solar cell design. We will focus on a non-silicon device platform that utilizes large surface area collection through nano-scale engineering – namely photoelectrochemical Gratzel cells.

This platform has the potential to become a serious low-cost, medium efficiency alternative to silicon. Furthermore, we will explore the possibilities of using bio-organic pigments with broad band light absorbing properties as the primary photon harvesting component. In this context, we will concentrate on a class of bio-macromolecules called the melanins, and will also highlight several other potential material. What will hopefully become clear during this discussion, is the need to understand the detailed photophysical and photochemical properties of these molecules, and how these properties relate to their chemical and electronic structure. It is also important to note that we will not deal in this chapter with the concept of creating artificial photosynthetic membranes by immobilizing and assembling photofunctional molecules on various substrates – although this approach has the potential to create direct conversion photovoltaic devices. The interested reader is directed towards a growing body of literature on the subject dealing with, for example, self-assembled monolayers, Langmuir-Blodgett membranes and cast films [1,2,3].

Finally, throughout this chapter we will adopt the specific definition that an **artificial photosynthetic system** for **photovoltaic energy** conversion is characterized by the following key features:

1. A large surface area for photon absorption created through molecular scale or nano-scale engineering.
2. Spatial decoupling of the primary photon absorption event from the charge separation event.
3. A photo-induced excited state sufficiently long lived to allow charge separation to occur without significant quenching from wasteful back reactions and recombination events.
4. Strong and rapid coupling between the absorber (or the charge separation vehicle in any harvesting chain system) and the electronic sink used to connect the photo-active component to the external circuit.
5. Ultimate generation of a potential difference across a nano-structured thin membrane, which requires charge separation to be directional.
6. Photochemically stable components capable of multiple oxidation / reduction cycles (or multiple exciton generation and separation events in a solid-state semiconductor system).

These are similar criteria to the key elements of any efficient light-driven system that separates charge and stores or converts the resultant chemical energy [4].

## 2. The Photoelectrochemical Gratzel Cell (Dye Sensitized Solar Cell)

The direct conversion of solar energy into electrical energy (photovoltaics) is a persuasive concept. Ever since the discovery in 1839 of the photoelectric effect by Edmond Becquerel [5], scientists and engineers have been obsessed with the idea of tapping this seemingly "free" and plentiful energy source. Photovoltaics relies upon the fact that solar photons (UV, visible or infra-red) falling upon a semiconductor with a suitable band gap can create electron-hole pairs. This effect can ultimately lead to the creation of a potential difference at an interface between two regions within the semiconductor possessing different electron affinities (for example in a *pn* junction) – or indeed at an interface between two different materials, as in a heterojunction device. To date, photovoltaics has been dominated by devices in which the junctions are made from crystalline or amorphous semiconductors. Doped silicon devices remain the lead technology, but compound semiconductor devices made from III/V compounds have made a limited impact in areas such as high efficiency aerospace applications.

More recently, since the early 90s onwards, advances in the nano-structuring of crystalline semiconductors have opened up the possibility of creating very large internal surface areas in relatively thin films (<10μm) [6]. Titanium dioxide (titania), a wide band gap semiconductor ($E_g$ ~3.2eV), is a particularly interesting material since it is cheap, readily available and has a commendable environmental profile. Total surface areas in titania nano-structured films can exceed a massive 150m$^2$/g. There is significant activity in the materials physics and chemistry community focused on increasing porosity (surface area), whilst controlling pore size, geometry and orientation [7]. Additionally, as we shall see later in this chapter, from a device perspective (whether it be photovoltaic or some other optoelectronic platform), the electrical conductivity of the nano-crystalline semiconductor network is a key consideration. In the case of titania for example, the anatase crystalline phase has the highest electrical conductivity and so is most desirable. Titania can also form two other crystal types – rutile and brookite, both of which suppress electrical conductivity of the network. Clearly, if one could create an interpenetrating mesoporous network of *n* and *p* type nanocrystalline titania (or of nanocrystalline titania and a different semiconductor / alternative hole transport medium), this would essentially correspond to a large number of nano-junctions for the absorption of light and interface driven exciton separation, i.e. a very high surface area and potentially very efficient solar cells.

However, there are significant difficulties associated with this Eutopian view of a nano-structured photovoltaic system. Firstly, it is extremely difficult to create a nano-structured interpenetrating network of two materials. Secondly, titania (and many of the other nano-crystalline semiconductors) are wide band gap materials – i.e. they only absorb UV photons. Narrower band gap semiconductors tend to be susceptible to photocorrosion. This is a serious draw back since the majority of the sun's energy is delivered as visible light. Thirdly, both components have to form a percolated network throughout the film. Holes have to migrate through one component to a suitable cathode, and electrons though the other component to a suitable anode. Mobilities for each charge carrier type must be high, and opportunities for recombination events must be minimized. This is clearly a very difficult task considering that the film consists of two interpenetrating nano-structured networks with many interfaces and tortuous percolation pathways.

A solution to these problems was suggested by Michael Gratzel and his group in 1991 [8]. Inspired by natural photosynthetic systems, Gratzel demonstrated a photoelectrochemical device based upon decoupling the primary light absorption and charge separation events. Retaining the nano-crystalline concept, the design used a visible light absorbing pigment chemically bonded in a monolayer to the surface of the semiconductor. This device became known as a nano-crystalline Dye Sensitized Solar Cell (DSSC). The "sensitizing component" was chosen such that, in the photo-excited state, it was capable of injecting an electron into the conduction band of its semiconductor host. Additionally, Gratzel solved the interpenetration problem by using a liquid electrolyte containing a redox couple as the hole transport medium. The theory of semiconductor-electrolyte interfaces is well established –a clear and elegant summary of the energetics is given in reference [5]. Conventional thought would suggest that a potential difference is established at such an interface. The nature of this interface potential depends very much upon the details of the electrolyte solution (specifically its redox potential) and the Fermi energy of the electrons in the solid. The surface of the semiconductor can be *p*-type or *n*-type dependent upon whether a depletion or accumulation layer is formed upon contact – i.e. whether the electrolyte preferentially accepts or donates electrons. Both situations can be manufactured by including electron or hole scavenging components in the electrolyte. The creation of a suitable contact potential would certainly aid the separation of a bound electron-hole pair at the "molecular" junction formed by the semiconductor and photo-excited sensitizer. However, this simplistic picture is complicated by the fact that nano-structuring of the semiconductor has a profound effect upon its photoelectrochemical properties. Critically, a depletion (or indeed accumulation) layer cannot be formed in the solid since the nano-crystallites are tiny (of the order of 10s of nms). The voltage drop within the particles remains very small and a significant local electric field cannot be maintained. Hence, it would seem that the photoresponse of such a system is dependent upon the relative rates of reactions of the positive and negative charge carriers with the electrolyte couple. Confirmation of this theory was provided by the demonstration of a switch between anodic and cathodic photo-response simply by changing the hole / electron scavenging nature of the electrolyte [5].

Irrespective of whether a local electric field aids the separation of charge, several facts are very clear. Firstly, the primary light absorption and charge separation events are decoupled in a DSSC. Secondly, the sensitizer provides a means of harvesting visible as well as UV solar photons. Thirdly, the liquid electrolyte is capable of fully percolating the nano-crystallite network, and provides an effective medium for hole transport to the cathode. Hence, it would seem that the Gratzel cell design is an elegant solution to the problems associated with creating a large surface area for photon absorption and charge separation. Additionally, according to the criteria outlined in the introduction, the nano-crystallite dye sensitized solar cell can clearly be thought of as an artifical photosynthetic system.

## 3. Typical Components and Performance of a DSSC

### 3.1 Construction and Mode of Operation

The schematic in Fig.1 shows a typical DSSC design based upon nano-crystalline $TiO_2$. The support substrate is normally glass [5], although it is perfectly possible to use a flexible plastic substrate. Light is coupled into the cell through this side, and so the support substrate must be transparent in the visible and near UV. A thin film of a transparent, conducting material is vacuum deposited on the inside of the support substrate and acts as the anode electrode. Semiconductors such as indium tin oxide (ITO) or fluorine doped tin oxide ($SnO_2$:F) are normally used for this purpose. A porous film of the nano-crystalline semiconductor forms the actual photoanode. These films are usually between 1 and 10 microns thick, and can be fabricated using a number of different techniques. The most popular method is to cast a slurry of the nano-crystals using spray, dip, spin or drag coating, and then calcine the film at 400˚- 450˚C to confer structural stability and create a percolated network. P25 is a nano-crystalline product sold by BASF. It has a high anatase fraction (>80%), and depending upon preparation of the colloid, contains 10-25nm nano-crystallites. We have found that P25 films made using a standard method [9], have total surface areas of ~50m$^2$/g, and average pore diameters of 50-100nm [10]. Several other hydrothermal methods can be used to produce TiO$_2$ of varying particle size and crystallinity [11]. Of particular interest is the microwave method of Wilson *et. al.*, [12] which produces particularly small crystallites (4-5nm) with a high anatase fraction.

An alternative approach to producing highly porous nano-structured titania photoanodes is via surfactant templating of a titania precursor, for example an alkoxide such as titanium isopropoxide or titanium chloride templated with the pluronic surfactant P123 [7]. This sol-gel technique relies upon controlled hydrolysis and condensation of the precursor to create a structured network of titania, followed by removal of the template and calcination at temperatures exceeding 300˚C. Using this method, highly ordered mesophases can be created with well controlled pore dimensions and geometries [7]. The films produced via this method have a completely different nano-morphology to those created using slurries of nano-crystals. In the former case, the titania forms the pore walls, whilst in the latter, the pores are actually the nano-crystal interstices. Fig.2 shows a set of transmission electron micrographs highlighting the morphological differences. Although the nano-crystal slurry method is rather "quick and dirty", it does seem to produce higher efficiency photo-anodes, and is the technique of choice in most laboratories. This is likely due to better uptake of the sensitizer because of the larger pores, and a higher degree of crystallinity. The templated mesoporous materials tend to contain some amorphous titania which hinders conduction. A relatively new and rather interesting approach is to combine the two methods in a two step process: firstly, create extremely small (<5nm) nano-crystals of anatase, and secondly, self assemble these nanocrystals around a suitable surfactant template to create an ordered mesophase. This approach may ultimately lead to better control over morphology and crystallinity, and hence to improved engineering of the photoanode. We have recently adopted this approach with encouraging early results [13].

Irrespective of the exact morphology and crystallinity of the photo-anode, the sensitizer must be adsorbed to its surface in a monolayer [5]. A significant amount of attention has been directed towards optimizing the sensitizer over the last decade or so [4]. The three key design parameters are:

1. The spectral absorption of the sensitizer: ideally, the material should be "black" in the UVA, visible and near IR, with an extremely high, broad band absorption.

2. The energetics and dynamics of the coupling between the photo-excited state of the sensitizer and the conduction band of the photo-anode: the transfer of an electron between the photo-excited sensitizer and the photo-anode must be rapid (when compared to the lifetime of any other relaxation or quenching processes) and energetically favorable.
3. The photostability of the sensitizer: the material must be capable of many reduction / oxidation cycles without decomposition.

The best photovoltaic performances (power conversion efficiency and stability) have been achieved with polypyridyl complexes of ruthenium or osmium. The N3 ruthenium complex *cis*-RuL$_2$(NCS)$_2$, where L=2,2'-bipyridyl-4-4'-dicarboxylic acid as shown in Fig.3, is the standard sensitizer, only surpassed in performance by "ruthenium black" (tri(cyanato)-2,2',2"-terpyridyl-4,4',4"-tricarboxylate RuII) which holds the world record power conversion efficiency for a DSSC (10.4% under AM1.5 global sunlight) [14]. The fully protonated N3 dye has absorption maxima at 380nm and 518nm, with respective extinction coefficients of 1.33 and 1.3x10$^4$M$^{-1}$cm$^{-1}$. The dye is anchored to the titania surface through carboxylate ligands, and the main optical transitions have a metal-to-ligand charge transfer character (MLCT): an excited electron is transferred from the metal centre to the $\pi^*$ system of the carboxylate ligand. This electron is subsequently injected into the titania conduction band within femto to picoseconds, producing charge with near unity quantum yield. If allowed to relax radiatively, the complex emits at 750nm with a characteristic lifetime of 60ns. Hence, the conduction band injection process is by far the most rapid process.

In order for the injection process to also be energetically favorable, the potential of the excited state electron in the sensitizer must be higher than the conduction band edge of the titania [15]. Fig.4 shows these parameters for N3 adsorbed onto undoped titania with a band gap of 3.2eV. Also shown in Fig.4 is the redox potential of the electrolyte containing a typical iodine/iodide couple. The role of the electrolyte is to "transport" the hole that is generated by charge separation to the cathode. Stated another way, the electrolyte must re-reduce (donate an electron from the redox couple) the oxidized sensitizer molecule in order to regenerate the system for another cycle. The electrolyte itself is reduced at the counter electrode, usually a low work function metal such as platinum, in order to complete the circuit. Critically, the redox potential of the couple must be higher than the ground state potential of the sensitizer. The difference between the quasi-Fermi level potential of the semiconductor and the redox potential of the electrolyte defines the maximum open circuit voltage that can be generated by the cell. It's also worth noting that this type of cell is a majority carrier device, since the hole (although we describe it as being "transported" to the cathode) remains localized on the oxidized dye molecule. The overall electric field that will tend to draw electrons to the anode, and holes to the cathode, is generated by the difference in work functions between the transparent electrode (ITO or $SnO_2$:F) and the metallic counter electrode. This is typically of the order of ~0.5 to 0.6 volts.

Summarizing the mode of operation of a typical DSSC such as the one shown in Fig.1:
1. A photon is absorbed by the sensitizer – the photo-anode (nano-structured titania plus sensitizer) presents a very large surface area for photon absorption.
2. The photo-excited sensitizer injects an electron into the conduction band of the titania within femto to picoseconds of the primary absorption event.
3. The injected electron will diffuse through the nano-structured titania and be delivered into the external circuit by the transparent anode electrode.
4. The oxidized dye is reduced by donation of an electron from the redox couple, and the electrolyte is in turn reduced at the counter electrode as the circuit is completed.

For a more comprehensive discussion of the energetics and thermodynamics of light induced redox reactions in such systems, the interested reader is directed towards reference [15].

### 3.2 Typical DSSC Performance

The performance of a solar cell is defined by a number of key parameters:
1. The short circuit current ($I_{sc}$) and open circuit voltage ($V_{oc}$) generated under standard illumination conditions (global AM1.5).
2. The Fill Factor (*FF*) of the cell under AM1.5 illumination. This quantity is obtained by a full current-voltage characterisation, and is essentially a measure of the diode behavior of the cell. The *FF* is given by:

$$FF = \frac{P_{max}}{V_{oc}I_{sc}} = \frac{(VI)_{max}}{V_{oc}I_{sc}} \qquad [1]$$

3. The Incident Photon Conversion Efficiency (*IPCE*), this is an incident energy dependent quantity, and is a measure of the useful range of the cell. The *IPCE* at a given illumination wavelength λ, under incident optical power $P_{in}$ is given by:

$$IPCE = \frac{P_{in}}{I_{sc}} \cdot \frac{e\lambda}{hc} \qquad [2]$$

Where *e* is the fundamental electronic charge, *h* is Planck's constant and *c* is the speed of light in vacuum.

4. The global power conversion efficiency (η) of a cell producing $P_{out}$ electrical power under standard AM1.5 illumination conditions given by:

$$\eta = \frac{P_{out}}{P_{in}} = I_{sc} V_{oc} \frac{FF}{P_{in}} \qquad [3]$$

Under full sunlight conditions (simulated by the AM1.5 standard illumination), with an incident radiant flux of 100mWcm$^{-2}$, the best nano-crystalline DSSC based upon N3 or a similar ruthenium based dye, produce short circuit current densities of 16-22mAcm$^{-2}$, and open circuit voltages of 0.65-0.75V. With fill factors of 0.65-0.75, these figures correspond to a maximum global power conversion efficiency of ~10.4%. Many laboratories report respectable efficiencies of 5-8%. Although these figures are fully competitive with the better amorphous silicon devices, the DSSCs are limited by dye liftetimes. In order to be a credible commercial alternative, the dye molecule must sustain at least 10$^8$ redox cycles to give a device working life of 20 years. With the current ruthenium based systems, this does not appear to be possible, although the use of purified solvents such as γ-butyrolactone has produced devices capable of passing standard quality assurance tests for outdoor use [16].

Fig.5 shows a typical normalised *IPCE* profile obtained in our laboratory using the N3 dye in combination with a templated hydrothermal seed nano-crystalline TiO$_2$ photoanode. The electrolyte for this cell contained 0.5M LiI and 0.04M I$_2$ as the redox couple, in 0.05M 4-*ter*-butylpyridine and acetonitrile. Clearly, the N3 dye is capable of producing a photoresponse from ~400nm to ~650nm, i.e. spanning a good proportion of the visible spectrum. The I-V characteristics of this cell are shown in Fig.6, and we see an open circuit voltage of 0.717V, short circuit current density of 12.12mAcm$^{-2}$ and fill factor of 0.58, yielding a respectable power conversion efficiency of ~5% under AM1.5 illumination of ~150mWcm$^{-2}$. No corrections to this data were made for factors such as back and front surface reflection losses or back surface leakage.

### 3.3 Device Limitations

Currently DSSCs are limited by a number of factors:

1. The lifetime (stability) of the sensitizer.
2. The low electrical conductivity of the nano-crystalline photo-anode.
3. The stability of the electrolyte system.
4. The overall stability of the cell, given that it contains a liquid component capable of leaking and evaporating if not completely sealed.

Numerous approaches have been suggested to address each of these limitations. Notably, replacement of the liquid electrolyte with a solid hole transport or ionic medium would solve the stability and leakage related issues. For example, cells based upon a spirobisfluorene-connected arylamine hole transmitter have produced a full sunlight efficiency of 2.56% [17]. Likewise, systems incorporating inorganic *p*-type semiconductors such as CuI or CuSCN have yielded 1% power conversion efficiencies [18]. In the inorganic case, difficulties associated with semiconductor photostability, and in creating good electrical contact with the nano-crystalline $TiO_2$, will probably limit the use of such materials. Currently, organic hole conductors, which can be applied as liquids and then cured into solids, would seem the most likely candidates to replace the liquid electrolyte [19].

The way forward with respect to increasing photoanode conductivity probably lies in better control of the film nano-morphology. The optimum system is likely to be one in which 100% anatase nano-crystals are assembled into pseudo 1D nano-structures orientated perpendicular to the support substrate. This would favor electron transport to the anode by minimizing lateral percolation pathways. In principle, it is already possible to build $TiO_2$ "micro-pillars" [20], and if this approach could be extended into the nano-regime in order to maintain high internal surface areas, it may be possible to build an orientated photo-anode for DSSC applications. An alternative approach, and one that is more realistic in the short term, is to create morphology control with templating. It appears as though the type of mesoporous networks that can be routinely produced using precursors such as titanium isopropoxide and templates such as block copolymer surfactants, are not suitable for DSSCs [7] even though they have surface areas exceeding $150 m^2 g^{-1}$. This is likely due to the reduced pore size (relative to the random nano-crystal films), and the "closed" nature of the mesophase – i.e. the sensitizer molecules cannot penetrate the structure. A possible solution to these issues would be to manufacture a meso-structured film with lamellar pores orientated perpendicular to the substrate, and pore-walls of pure nano-crystalline anatase. To the author's knowledge, such a film has yet to be fabricated, but with the rapid advances being made in molecular self-assembly, it may be possible in the near future. Finally, a two stage anatase nano-crystal templating process as discussed in section 3.2 may be a realistic near term solution. If tiny "seed" crystals can be assembled around a block copolymer template, and the system subsequently calcined to remove the surfactant and form the percolated nano-crystal network, then better morphology control would undoubtedly lead to improved photoanode conductivity and hence device performance.

The ruthenium N3 and ruthenium black dyes are the best sensitizers (in terms of efficiency) demonstrated so far. They have relatively broad absorption profiles, display very rapid injection dynamics, have long lived excited states and can be engineered to adsorb strongly to titania. On the down side, they are expensive to manufacture, do not display the desired long term stability and do not have the ability to harvest near infrared photons. Several alternatives to the ruthenium or osmium MLCT complexes have been suggested [21,22,23]. The next section will deal with one material in particular – a family of broad band absorbing and ultra stable bio-macromolecular pigments called the melanins.

## 4. Melanins as Broad Band Sensitizers for DSSCs

### 4.1 Melanin Basics

The melanins are an important class of pigmentary macromolecules found throughout nature [24]. Eumelanin is the predominant form in humans, and acts as the primary photoprotectant in our skin and eyes. Unique amongst bio-macromolecules, the melanins are broadband UV and visible light absorbers (Fig.7). Their spectra display no chromophoric bands and follow a smooth single exponential dependence on wavelength increasing monotonically towards the UV [25]. Most melanins are also chemically and photochemically very stable, and are potent free radical scavengers and antioxidants [24]. In direct contradiction with its photoprotective properties, eumelanin (along with pheomelanin – the less prevalent red-brown pigment also found in humans), is implicated in the cytotoxic chain of events that ultimately lead to melanoma skin cancer [26]. For this reason, the photophysics, photochemistry and photobiology of melanins are subjects of intense scientific interest.

In a more general sense, the broader structure-property-function relationships that dictate the behavior of these important biological macromolecules are still poorly understood [27]. In particular, major questions still remain concerning the basic structural unit [28]. It is fairly well accepted that eumelanins are macromolecules of 5,6-dihydroxyindole (DHI) and 5,6-dihydroxyindole-2-carboxylic acid (DHICA) (Fig.8), and that pheomelanins are cysteinyl-dopa derivatives [29,24]. However, it is still a matter of debate as to whether eumelanin (in particular) is actually a highly cross-linked extended heteropolymer, or composed of DHI/DHICA oligomers condensed into 4 or 5 oligomer nano-aggregates [30]. This is an absolutely fundamental issue, and is the starting point for the construction of consistent structure-property-function relationships. The answer to this question also has profound implications for our understanding of the condensed phase properties of melanins. In 1974, McGinness, Corry and Proctor showed that a pellet of dopamelanin could be made to behave as an amorphous electrical switch [31]. They postulated that these materials may be disordered organic semiconductors. Several studies since have claimed to show that melanins in the condensed solid-state are indeed semiconductors [32,33]. However, it is by no means certain that the conductivity reported is electronic in nature. A clear idea of the basic structural unit is fundamental to developing a consistent model for condensed phase charge transport in such disordered organic systems.

From the preceding discussion it is clear that melanins may have some potential as photon harvesting systems. Their broad band absorbance and photostability are key attributes, as are their bio-compatibility and bio-availability. However, as we saw in section 3, sensitizer excited state energetics and compatibility with a suitable nano-stuctured semiconducting host are also key considerations in DSSC applications. Work in our laboratory over the past few years has concentrated on understanding these important properties with respect to natural melanins and tailored synthetic analogs. This application focused work is part of a broader program of condensed matter physics, synthetic organic chemistry, quantum chemistry and molecular biophysics aimed at gaining a more complete understanding of melanin structure-property-function relationships. In the following sections, we outline some recent results pertinent to the photon harvesting concept, and ultimately, demonstrate that it is possible to couple these materials to a nano-crystalline $TiO_2$ host in order to produce visible light sensitization and hence a regenerative DSSC system.

## 4.2 Melanin Chemical, Structural & Spectroscopic Properties

As highlighted above, melanins are broad band absorbers with a monotonic exponential dependence of extinction upon wavelength. Fig.7 shows typical absorption spectra for a synthetic dihydroxyindole eumelanin in aqueous solution. Fig. 9 summarizes the absorption coefficient values at 380nm for the three concentrations of Fig.7. It is instructive to use this data to calculate molar extinction coefficients and to compare with a typical ruthenium sensitizer such as N3. Unfortunately, due to the difficulties associated with determining the molecular weight of the eumelanin sample in solution [34], a molar based comparison cannot be performed. However, it is possible to calculate a weight-based extinction coefficient, and this comparison (Table 1) shows that the synthetic eumelanin is a more effective absorber gram for gram than N3 at both UV and visible wavelengths.

Melanins have also been shown to exhibit some rather exotic emission and excitation as well as absorption behavior. Several steady-state photoluminescence studies have demonstrated that melanins possess an extremely low radiative quantum yield (i.e. the fraction of excitation photons that ultimately lead to a spontaneously emitted photon) [35]. We have recently reported the first accurate measurement of yield for a synthetic eumelanin, and found it to be $<6 \times 10^{-4}$ i.e. 99.94% of absorbed photon energy is dissipated non-radiatively [34]. This measurement is particularly difficult for a number of reasons: firstly, the emission is extremely weak (as evidenced by the very low quantum yield), and secondly, melanins have strong broad band absorption – a fact that leads to heavy emission re-absorption and probe beam attenuation. We have developed a method for accounting for these factors, and hence have been able to properly quantify the photoluminescence emission spectra. Fig. 10A and fig.10B show a series of these spectra for three aqueous solutions of eumelanin (0.001, 0.0025 and 0.005% by weight). Fig. 10A contains the raw data – heavily affected by probe beam attenuation and emission re-absorption (as evidenced by the inconsistent peak position and lack of correspondence between peak height and concentration). Fig. 10B and fig. 10C show how the expected linear relationship between emission and concentration is recovered after re-correction, as is the consistency in peak position. In general, the photoluminescence is characterized by a broad, single peak with a red shift of ~100nm relative to the excitation wavelength.

Eumelanin photoluminescence shows a number of other unusual features. Fig. 11 demonstrates how the emission is strongly dependent upon excitation energy (wavelength). Peak position and height change as a function of pump energy, with the red edge appearing to collapse towards a fixed value. Surprisingly, the radiative quantum yield (fig. 12) also shows strong excitation dependence – an extremely unusual property amongst organic chromophores. We believe that our observations are consistent with an ensemble model: many chemically distinct species existing within the system, each with a different HOMO-LUMO gap (difference in energy between the highest occupied and lowest unoccupied molecular orbitals), emission characteristics and radiative quantum yield. The broad emission is a result of the overlap of individual emission peaks, and the excitation energy dependence arises due to selective pumping of a particular sub-set of the ensemble. However, these chemically distinct species could only produce different optical signatures if they were small (i.e. non-polymeric molecules). Hence, our data lends further credence to the argument that the basic structural unit of melanins (eumelanin in particular) is oligomeric rather than hetero-polymeric in nature. This view is also supported by time resolved photoluminescence emission studies on eumelanin solutions, which consistently show non-exponential behavior, and can only reconciled by fitting the data to a distribution of lifetimes (fig. 13). The ensemble model could also explain the broad band absorption of melanins. The distribution of HOMO-LUMO gaps (each manifesting itself as an inhomogeneously broadened Gaussian feature in the absorption spectra) could overlap to produce a single monotonic profile.

The above argument clearly relies upon the fact that small macromolecules can form, and that furthermore, these distinct macromolecules have different fundamental absorptions (HOMO-LUMO gaps). Hence, we are once again forced to return to the vexing question of melanin fundamental structure. In the case of eumelanins, the macromolecular entities are based upon DHI and DHICA. It is well known that these two monomers exist in various redox forms, namely the indolequoinone (IQ), semiquinone (SQ) and hydroquinone (HQ) [36]. These structures are shown in fig. 14. Quantum chemical methods have proven somewhat useful in predicting the electronic and vibronic properties of these monomeric redox forms, and the macromolecules that can be assembled from them. The first major attempt to study melanins in this way was due to Longuet-Higgins [37], who proposed that the optical (and indeed electronic) properties could be explained if the macromolecule was an infinite homopolymer - an assumption that we now know to be incorrect. Later work used semi-empirical methods [38] such as Hückel theory [39,40,41,42] and the intermediate neglect of differential overlap (INDO) [43,44]. These techniques were applied up to the small oligomer level, but crucially, the simulated spectra did not show the expected broad band absorption.

More recently we (and others) have applied *ab initio* methods, particularly density functional theory (DFT) [45], to both DHI [27,36,46] and DHICA. In DFT one maps the Schrodinger equation onto a non-interacting equation for the electronic density, known as the Kohn-Sham equation. While it can be shown that this mapping gives (at least in principle) the exact ground state energy, in practice one important term called the exchange correlation function ($E_{xc}$) must be approximated. There are basically two approaches to approximating $E_{xc}$: either the fact that DFT is in principle exact is used to derive $E_{xc}$ (the current state-of-the-art functional of this type is due to Perdew, Burke and Ernzerhof (PBE) [47]), or else $E_{xc}$ is extracted empirically from a large number of experimental results (for example the hybrid functional B3LYP [48]). Both types of functional have been successfully applied to study the building blocks of eumelanin. One major drawback with DFT is that, as it is a theory of the ground state, excited state properties are not given accurately by the simple interpretation of the Kohn-Sham eigenvalues. This is known as the band gap problem [45]. The optical absorption of a molecule is essentially an excited state property. Both time dependent DFT (TDDFT) calculations and the difference of self consistent fields ($\Delta$SCF) method have been applied to DHI and DHICA to avoid the band gap problem. However, the functionals are not as well developed for TDDFT as for standard DFT. In the $\Delta$SCF method one calculates the total energy of the ground state and the total energy of an excited state (in the ground state geometry). One then takes the difference of these two energies to be the energy gap to the excited state. Although, except in a few special cases, there is no mathematical justification for this approach, in practice it works rather well [36,45].

In Table 2, we compare the results from the various published calculations. It can be seen that the trends found by all of the methods are in broad agreement. More importantly, the quantitative results of the *ab initio* calculations (TDDFT and $\Delta$SCF) are in excellent agreement. Critically, these results indicate that the HOMO-LUMO gaps of the DHI redox forms (IQ, SQ and HQ) are appreciably different. In addition, it has been recently demonstrated that several redox forms and charge states of DHI [27,30,36,44] and DHICA are thermodynamically stable. Hence, given that the basic monomer can cross-link in a number of possible positions (2, 3, 4 and 7) we are led to conclude that a wide variety of "chemically distinct" oligomers, each with a different HOMO-LUMO gap, can form. As the HOMO-LUMO gap is closely related to the fundamental absorption, it is reasonable to expect that absorption spectra derived from such an ensemble of monomers and oligomers would contain a large number of overlapping optical absorption peaks. In the limiting case, if the system contained enough of this "chemical disorder", then the individual peaks would become smeared into a monotonic absorption, or a single broad emission feature in the case of photoluminescence.

Presently, this disordered ensemble model is no more than a hypothesis. More advanced DFT calculations are underway to predict the stability and HOMO-LUMO gaps of larger eumelanin oligomers that result from the cross-linking of DHI and DHICA redox forms and charged states. We are also extending this study to pheomelanin (a cysteinyldopa derived melanin), and other synthetic eumelanins derived from protected forms of DHI and DHICA.

Once validated by direct comparison with experiment, this new structural information will be useful from a number of perspectives: firstly, it will form the basis of mesoscopic models to explain phenomena such as electrical conductivity and photoconductivity, secondly, it will guide molecular engineering strategies as we seek to optimize functionality for a particular application (for example photovoltaics), and thirdly, it will form an integral part of broader attempts to explain melanin structure-property-function relationships in biology. With respect to the last point, it is an interesting fact (especially in light of the subject matter of this book – Artificial Photosynthesis) that biomimetic strategies often concentrate on imitating the exquisite structural order that pervades the natural world. In a general sense, many biological macromolecules derive their functionality from precise and directed ordering at several structural levels. Structural biologists and molecular biophysicists are pre-programmed into looking for this directed structure to explain properties and function. If the disordered ensemble model of melanins is correct, the structure-property-function relationships of these materials could represent a significant paradigm shift – functional utility, flexibility and robustness are derived from chemical and structural *disorder* rather than *orde*r. It's a common mantra in melanin biophysics that no two melanin molecules are ever made the same. In the context of the disorder model, there may well be a very good reason for this fact.

**4.3 Melanin Electrical and Photoconductive Properties**

In 1974, McGiness *et al*. [31] showed that a solid pellet of eumelanin (the material is normally a powder when extracted from pigment containing tissues or when synthesized chemically) could conduct electricity. At applied electric fields of ~350Vcm$^{-1}$, the material was observed to switch between "low" (~$10^{-5}$Scm$^{-1}$) and "high" ($10^{-3}$Scm$^{-1}$) conducting states. This behavior led the authors to postulate that eumelanin was acting as an amorphous semiconductor – a hypothesis originally advanced in a landmark theoretical paper by Longuet-Higgins in 1960 [37]. The amorphous semiconductor model of eumelanin is now relatively widely accepted. A number of experimental studies have claimed not only to confirm this viewpoint, but also to derive band gaps, activation energies, carrier types and densities, etc. using DC and AC conductivity measurements, thermopower, photoconductivity and optical absorption in combination with standard Mott-Davies theory [32,49]. What is worrying about many of these studies, is the lack of any serious consideration of the effect of adsorbed water – as demonstrated by Jastrzebska *et al.*, eumelanin conductivity is highly dependent upon atmospheric relative humidity [50]. The latter authors also claim to show some temperature dependence of the conductivity (σ), but only make measurements over a very limited range (290K to 340K in-vacuum). They go on to fit the data using the standard thermally activated semiconductor model (equation 4), and extract an activation energy ($E_a$). This is clearly inadequate – normally, one requires at least two to three orders of magnitude in temperature before drawing such conclusions.

$$\sigma = \sigma_0 \, exp(-\Delta E_a / kT) \qquad [4]$$

A general model for electrical conductivity in condensed phase eumelanin remains elusive. Decoupling the effect of adsorbed water requires high vacuum, and to extract any meaningful information from DC conductivity vs. temperature data requires measurements over several orders of magnitude. Fig.15 shows the conductivity of a pressed pellet of synthetic eumelanin (derived from DC current-voltage measurements) as a function of relative humidity at room temperature. It shows an extremely strong dependence – greater than 5 orders of magnitude in 80%. Apart from suggesting that these pellets may be very sensitive relative humidity sensors, it confirms the dominating nature of adsorbed water [51]. Measurements under vacuum as a function of temperature produce dramatically different results. At $10^{-6}$ torr, even below 10K, we observed the samples to be too insulating for a conventional I-V system to make a meaningful measurement of current at applied voltages of <20V. Consequently, we can say that the resistance of the sample was >1GΩ, with corresponding electrical conductivities of $<10^{-9}$ Scm$^{-1}$. Clearly, any electronic contribution to the conductivity (which would have been evident under these conditions), is extremely small.

Considering Fig.15 and the vacuum conductivity measurements in isolation, one might be led to conclude that eumelanin is essentially an insulator in the condensed solid state, and that all the conductivity was derived from ionic sources (adsorbed water). However, there are several reports [50] of solid films or pellets of eumelanin producing a significant photocurrent under white light illumination. Fig. 16 shows the photocurrent produced by a synthetic eumelanin film produced by the electropolymerisation of dopa [52]. These films are structurally more continuous than the pressed powder pellets, and display higher room temperature electrical conductivities (although from the preceding discussion, this is likely due to water absorption capacity). The photocurrent was produced by dropping a 20V potential across electrodes (~3mm apart) deposited on the film surface. Illumination was provided by an Hg vapor lamp (150W). Under these conditions a photocurrent of ~0.3μA was generated. The effect was stable and repeatable. The film showed capacitive recovery after the lamp was switched off, consistent with a semi-insulating material. Additionally, illumination caused a heating of the material and subsequent loss of bound water. This was shown by the fact that the instantaneous resistance of the film after illumination was higher than before illumination, but recovered to the original equilibrium value after several minutes.

In general, the production of a photocurrent under white light illumination is indicative of some "*semiconductor-like*" behavior. Ultra violet and visible light stimulates a significant number of charge carriers – an observation that can only be reconciled with a band gap (at least of one component within the system) of less than ~3.5eV. Hence, given the conductivity and photoconductivity data, we are forced to consider a hybrid model for charge transport in eumelanins (and likely all other melanins in the condensed solid state): small semiconducting regions (grains) containing delocalized electrons, coupled together by an essentially ionic (and dominating) percolated medium of hydrated amorphous material. It does appear as though the amorphous semiconductor theory first proposed by Longuett-Higgins [37], and supported by McGiness *et al*. [31] and many others since [49], may be an oversimplification of a much more complicated hybrid situation. More detailed electrical studies, and a predictive mesoscopic model based upon an accurate description of the melanin basic structural unit are required in order to advance this hypothesis.

## 4.4 Melanins as Broad Band Photon Harvesting Systems

Returning to the key motivation of this chapter, i.e. biomaterials as broad band photon harvesting systems, in way of summary, we now ask the two key questions: have the melanins got what it takes to be photovoltaic materials, and can they be integrated into a suitable device platform?

Addressing the first question: Clearly, melanins (eumelanin in particular) have strong, broad band absorbance - comparable with a typical sensistizer such as N3. At least some components within the "eumelanin" ensemble have the potential to absorb solar radiation (UV and visible), and generate photo-excited charge carriers. These are both positives from the photovoltaic perspective, as are the aforementioned attributes of bio-compatibility, bio-availability, ease of synthesis, chemical and photochemical stability (capability to cycle through many redox cycles), and general "green credentials". On the down side, the extremely low radiative quantum yield is indicative of strong excited state-phonon coupling. Thermal relaxation processes are characteristically very rapid (ps). Ideally (referring back to the introduction), a long excited state lifetime in the photon harvesting molecule is desirable in order to promote charge separation. However, the origins of the non-radiative relaxation modes in melanins are by no means clear. Paradoxically, any radiative decay is characterized by ns lifetimes – an encouraging fact from the photovoltaic perspective. Clearly, we need to understand more about the energy dissipation pathways in these systems before attempting to engineer a long excited state lifetime. Finally, with reference to the charge transport in eumelanins, we have seen electrical conductivity is dominated by adsorbed water and not by electronic processes. Hence, one could conclude that, in a solid thin film of these materials, the mean free path of any delocalized electron or hole would be very limited indeed. If one takes the oligomeric nano-aggregate model as a realistic description, the mean free path would likely correspond to a single oligomer unit, i.e. a few nms. It should be noted that no attempts have yet been made to dope these systems. So-called "inherently conducting polymers" display similarly low electrical conductivities, poor mobilities and mean free paths in their undoped states.

Addressing the second question: On the positive side, it is possible to produce tailored synthetic analogues of eumelanins (and other melanins) in order to maximize adsorption to, and coupling with a substrate surface. For example DHICA macromolecules can be synthesized with a high degree of carboxylation in the 2 position. This is a favored ligand for adsorbing and binding ruthenium sensitizers to titania in the DSSC platform. It is also possible to electropolymerise synthetic eumelanins into the pores of nanocrystalline titania. However, sufficient control has not yet been achieved to produce a monolayer of the pigment. Chemisorption of the macromolecule onto nanoporous titania tends to produce "pore blocking" which inhibits electrolyte ingression and efficient coupling of the photo-excited molecule to the titania host.

**4.5 A DSSC Based Upon Synthetic Eumelanin**

So, given what we now know about the physics and chemistry of melanins, and given the criteria for sensitization of a narrow band gap nanocrystalline photo-anode (titania for example), is it possible to use eumelanin as the light harvesting component in a DSSC? Critically, for such a system to function as a regenerative photovoltaic device, the photo-excited state of the eumelanin molecule (whatever that is) must be strongly coupled to the semiconductor conduction band.

We have recently fabricated a DSSC based upon nanocrystalline P25 titania sensitized with a synthetic eumelanin. The eumelanin was electropolymerised into the $TiO_2$ pores from a buffered aqueous solution of dopa [52]. To avoid any chance of over-deposition and hence pore blocking, the titania substrate was only partially sensitized – the resultant photo-anode was a light brown colour. Clearly this is non-optimum in terms of total light capture, but the object of the exercise was to determine whether or not the eumelanin was capable of injecting an excited electron into the conduction band of its host. As per Fig.1, $SnO_2$:F on glass was used as the transparent anode electrode and support substrate, an iodine / tri-iodide couple in acetonitrile was used as the redox electrolyte, and a thin film of platinum was used as the counter electrode (deposited on F:$SnO_2$ on glass). The cell was sealed with a silicone based polymer to prevent electrolyte leakage, and was tested versus a plain titania photo-anode cell under standard Air Mass 1.5 global illumination using an Oriel Solar Simulator. Current-voltage scans (I-V) were obtained using a Keithley SMU 2400 Source Meter Unit controlled using in-house computer code. Lamp output was measured using a NIST standard photodiode in order to establish the incident optical power density. Equations 1 and 3 were used to determine fill factors and power conversion efficiencies from the I-V data.

The I-V performance of the bare titania photo-anode cell and the eumelanin sensitized cell are shown in Fig.17. Under AM1.5 conditions, a relatively small amount of the incident light is ultra violet (as per the natural solar spectrum). The titania photo-anode cell shows the expected response. Titania itself produces a photo-current if illuminated with light above its band edge (~3.2eV). The eumelanin based cell produces a significantly higher short circuit current (photocurrent). This clearly indicates conversion of visible light into electrical energy, i.e. the eumelanin is sensitizing the titania, and electron injection is taking place.

This is an important result since it proves that coupling between the photo-excited state of eumelanin and the titania conduction band is possible. The fill factor of the cell was calculated to be ~0.4, and given a short circuit current of 0.81μA and open circuit voltage of 0.42V, this yields a global power conversion efficiency of ~0.1%. Clearly, this performance is a long way short of the 10-15% achieved with commercial silicon cells, and the 5% with a typical N3 DSSC. However, the eumelanin cell was far from optimum - in particular the photo-anode was only partially sensitized. These early prototype results are encouraging, and armed with the knowledge that eumelanin sensitization of titania is possible, we are attempting to optimize the cell for greater power conversion efficiency.

## 5. Conclusions

In this chapter we have discussed the criteria for classifying a photovoltaic system as artificially photosynthetic (AP). A key difference between semiconductor-junction based cells and AP systems is the removal of the need for spatial correspondence between the photon absorption and charge separation events. The photoelectrochemical Graztel cell (or dye sensitized solar cell) is a typical example of a biomimetic solar energy conversion system. These devices make use of the principles of artificial photosynthesis, and have competitive power conversion efficiencies relative to current silicon technology. Sensitizers based upon ruthenium charge transfer complexes (the N3 dye is a good example) produce a broad band UV and visible response. However, these complexes have long term stability problems and are expensive to manufacture.

We have suggested that broad band absorbing pigments based upon the melanins – a class of biological macromolecule, may be suitable replacements for the ruthenium complexes as sensitizers. Eumelanin in particular, is chemically and photochemically very stable, has strong broad band absorption, can be synthesized and engineered to maximize coupling and adsorption to a semiconductor host, is bio-compatible and even bio-available. However, before we can truly realize the potential of these molecules as photon harvesting systems, key questions regarding fundamental structure and mesoscopic physics need to be addressed. Despite these knowledge gaps, we have recently demonstrated a regenerative DSSC based upon an electropolymerised eumelanin photo-anode. Although the power conversion efficiency in this non-optimized device was modest, we have proven that it is possible to couple the photo-excited state of the eumelanin to the nanocrystalline titania conduction band. Hence, we have shown that these materials have great potential as broad band light harvesting components in biomimetic photovoltaics.

**Acknowledgements**

This work was funded by the Australian Research Council under Discovery grant DP0345309, and through the University of Queensland's Research Infrastructure Fund. We acknowledge the contribution of the Centre for Microscopy and Microanalysis at the University of Queensland for assistance in electron microscopy. We also acknowledge David Menzies from Monash University in Melbourne for advice concerning the N3 cell construction, and Dr. Adam Micolich at the University of New South Wales for low temperature electrical measurements. Finally, we would like to thank Prof. Ross McKenzie (Condensed Matter Theory Group, University of Queensland Physics Department), Prof. John Simon (Duke University, USA), Prof. Tad Sarna (Jagiellonian University, Poland), Dr. Mark Pederson (Naval Research Labs, USA), Dr. Tunna Baruah (NRL and Georgetown University) and Dr. John McGinness for fruitful discussion and guidance.

**References**

[1] H. Imahori, Y. Mori and Y. Matano, *J. Photochem. Photobiol. C: Photochem. Reviews*, **4**, 51-83 (2003).

[2] Y. Saga and H. Tamiaki, *J. Photochem. Photobiol. B: Biology*, **73**, 29-34 (2004).

[3] Y. Kureishi, H. Tamiaki, H. Shiraishi and K. Maruyama, *Bioelectrochem. Bioenerg.*, **48**, 95-100 (1999).

[4] M. Gratzel, *J. Photochem. Photobiol. C: Photochem. Reviews*, **4**, 145-153 (2003).

[5] M. Gratzel, *Nature*, **414**, 338-344 (2001).

[6] B. J. Scott, G. Wirnsberger, M. D. McGehee, B. F. Chmelka and G. D. Stucky, *Adv. Mater.,* **13**, 1231-1234 (2001).

[7] R. Vogel, P. Meredith, I. Kartini, M. Harvey, J. D. Riches, A. Bishop, N. Heckenberg, M. Trau and H. Rubinsztein-Dunlop, *ChemPhysChem*, **4**, 595-603 (2003).

[8] B. O'Regan and M. Gratzel, *Nature*, **353**, 737-740 (1991).

[9] M. Gratzel, *J. Am. Chem. Soc.*, **123**, 1613-1624 (1993).

[10] I. Kartini, P. Meredith, J.C. da Costa, J.D. Riches and G.Q. Lu, *Curr. Appl. Phys.*, **4**, 160-162 (2004).

[11] P. A. Venz, J. T. Kloprogge and R. L. Frost, *Langmuir*, **161(11)**, 4962-4968 (2000)

[12] G. J. Wilson, G. D. Will, R. L. Frost and S. A. Montgomery, *J. Mat. Chem.*, **12**, 1787-1791 (2002).

[13] I. Kartini, D. Menzies, D. Blake, J. C. D. da Costa, P. Meredith, J. Riches and G. Q. Lu, *submitted J. Mater. Chem.,* (2004).

[14] M. K. Nazeeruddin, P. Pechy, T. Renouard, S. M. Zakeeruddin, R. Humphrey-Baker, P. Comte, P. Liska, L. Cevey, E. Costa, V. Shklover, L. Spiccia, G. B. Deacon, C. A. Bignozzi and M. Gratzel, *J. Am. Chem. Soc.*, **123**, 1613-1624 (2001).

[15] A. Hagsfeldt and M. Gratzel, *Chem. Rev.,* **95**, 49-68 (1995).

[16] A. Hinsch, *Proc. 16th Eur. PV Solar Energy Conf.*, Glasgow, 32 (2000).

[17] U. Bach, D. Lupo, P. Comte, J. E. Moser, F. Weissortel, J. Salbeck, H. Spreitzer and M. Gratzel, *Nature*, **395**, 583-585 (1998).

[18] K. Tennakone, G. R. R. A. Kumara, A. R. Kumarasinghe, K. G. U. Wijayantha and P. M. A. Sirimanne, *Semicond. Sci. Technol.*, **10** 1689-1693 (1995).

[19] J. Kruger, U. Bach and M. Gratzel, A*ppl. Phys. Lett.,* **79**, 2085-2087 (2001).

[20] S. Z. Chu, K. Wada, S. Inoue and S. Todoroki, *Chem. Mater.*, **14**, 266-272 (2002).

[21] P. Meredith, PCT/AU02/01327: "*Components based on melanin and melanin-like biomolecules, and process for their production*", Filed September (2002).

[22] F. Odobel, E. Blart, M. Lagree, M. Villieras, H. Boujtitia, N. El Murr, S. Caromori and C. A. Bignozzi, *J. Mater. Chem.*, **13**, 502-510 (2003).

[24] G. Prota (1992) *Melanins and Melanogenesis*. Academic Press, San Diego, CA.

[23] G. Ramakrishna and H. N. Ghosh, *J. Phys. Chem. B*, **105**, 7000 (2001).

[25] M. L. Wolbarsht, A. W. Walsh and G. George, *Appl. Opt*. **20**, 2184-2186 (1981).

[26] H. Z. Hill, *Melanin: Its role in human photoprotection* (Edited by L. Zeise, M. Chedekel and T. Fitzpatrick), 81-91. Valdenmar Press, Overland Park, KS (1995).

[27] K. B. Stark, K. B., J. M. Gallas, G. W. Zajac, M. Eisner and J. T. Golab, *J. Phys. Chem. B* **107**, 3061-3067 (2003).

[28] Z. W. Zajac, J. M. Gallas, J. Cheng, M. Eisner, S. C. Moss and A. E. Alvarado-Swaisgood, *Biochim. Biophys. Acta,* **1199**, 271-278 (1994).

[29] S. Ito, *Biochim. Biophys. Acta,* **83**, 155-161 (1986).

[30] C. M. R. Clancy, J. B. Nofsinger, R. K. Hanks and J. D. Simon, *J. Phys. Chem. B* **104**, 7871-7873 (2000).

[31] J. McGinness, P. Corry and P. Proctor, *Science* **183**, 853-855 (1974).

[32] P. R. Crippa, V. Cristofoletti and N. Romeo, *Biochim. Biophys. Acta* **538**, 164-170 (1978).

[33] M. M. Jastrzebska, H. Isotalo, J. Paloheimo, H. Stubb and B. Pilawa (1996), *J. Biomater. Sci. Polymer Edn.* **7**, 781-793 (1996).

[34] P. Meredith and J. Riesz, *Photochem. Photobiol.*, **79(2)**, 211-216 (2004).

[35] J. B. Nofsinger and J. D. Simon (2001), *Photochem. Photobiol.* **74**, 31-37 (2001).

[36] B. J. Powell, T. Baruah, N. Bernstein, K. Brake, R.H. McKenzie, P. Meredith and M.R. Pederson, *J. Chem. Phys.,* **120(18)**, 8608-8615, (2004).

[37] H.C. Longuet-Higgins, *Arch. Biochim. Biophys.*, **88,** 231-232 (1960).

[38] J. P. Lowe, *Quantum Chemistry*, Academic Press, London (1978).

[39] A. Pullman and B. Pullman, *Biochim. Biophys. Acta,* **54**, 384 (1961).

[40] D. S. Galvao and M. J. Caldas, *J. Chem. Phys.*, **88**, 4088 (1988).

[41] D. S. Galvao and M. J. Caldas, *J. Chem. Phys.*, **92**, 2630 (1990).

[42] D. S. Galvao and M. J. Caldas, *J. Chem. Phys.*, **93**, 2848 (1990).

[43] L. E. Bolivar-Marinez, D. S. Galvao, and M. J. Caldas, *J. Phys. Chem. B*, **103**, 2993 (1999).

[44] K. Bochenek and E. Gudowska-Nowak, *Chem. Phys. Lett.*, **373**, 523 (2003).

[45] R. O. Jones and O. Gunnarsson, *Rev. Mod. Phys.*, **61**, 689 (1989).

[46] Y. V. Il ichev and J. D. Simon, *J. Phys. Chem., B*, **107**, 7162 (2003).

[47] J. P. Perdew, K. Burke, and M. Ernzerhof, *Phys. Rev. Lett.,* **77**, 3865 (1996).

[48] A. D. Becke, *J. Chem. Phys.,* **98,** 5648 (1993).

[49] T. Strzelecka, *Physiol. Chem. Phys.*, **14**, 219-222 (1982).

[50] M. Jastrzebska, A. Kocot and L. Tajber, *J. Photochem. Photobiol. B: Biology*, **66**, 201-206 (2002).

[51] P. Meredith, J. Riesz, C. Giacomantonio, S. Subianto, G. Will, A. Micolich and B. Powell, *International Congress on Synthetic Metals*, Wollongong (2004).

[52] S. Subianto, G. Will and P. Meredith, *International Congress on Synthetic Metals*, Wollongong (2004).

**List of Tables**

**Table 1** Comparison of extinction coefficients at 380nm and 518nm (weight-based) for the ruthenium N3 complex and synthetic eumelanin.

**Table 2** The calculated HOMO-LUMO gaps in eV for some key eumelanin monomers. The reduced forms of DHI and DHICA are indicated by bracketed numbers after the name. These numbers correspond to the location from which the hydrogen atom is removed. The functional used is listed in brackets after the name of the method. "Simple" implies a straightforward interpretation of the Kohn-Sham eigenvalues.

**List of Figures**

**Figure 1** Schematic representation of a typical nano-crystalline titania dye sensitized solar cell. Light is incident upon the device through the front transparent substrate and electrode (TCE = transparent conducting electrode). The sensitizer is bound to the surface of the nano-crystalline titania in a monolayer, and the whole photo-anode is percolated with an electrolyte containing a redox couple. The circuit is complete by a counter electrode (cathode) made from a thin film of a low work function material such as platinum.

**Figure 2** Bright field transmission electron microscopy (TEM) images showing the different nano-morphologies in nano-crystalline films made by hydrothermal (A) and mesoporous templating (B) techniques. The hydrothermal film contains individual nano-crystals sintered together to form a percolated network by calcination. The templated film (created from the hydrolysis and condensation of titanium isopropoxide and templated by the tri-block copolymer P123), contains an ordered mesophase of pores whose walls are composed of nano-crystalline titania. The inset in Fig. 2A contains an electron diffraction pattern with circular diffraction rings characteristic of a polycrystalline material as would be expected.

**Figure 3** The structure of the ruthenium complex N3 (adapted from reference [4]).

**Figure 4** A simplified energy level scheme for a DSSC based upon nano-crystalline titania and a ruthenium charge transfer complex as sensitizer. NHE refers to the standard hydrogen electrode reference.

**Figure 5** Incident photon conversion efficiency (*IPCE*) for a DSSC made from a templated hydrothermal seed nano-crystalline titania (Fig. 2A) sensitized with N3. The *IPC*E has been normalised at the peak response for clarity, and the graph shows two nano-crystal preparation methods [13].

**Figure 6** I-V response of the N3 cell from Fig. 5 under white light AM1.5 illumination (A) and in the dark (B).

**Figure 7** Absorption coefficient vs. wavelength in the near UV, visible and near IR of a synthetic eumelanin (aqueous solution) at 3 different concentrations (0.005% by weight - dotted, 0.0025% by weight - dashed, 0.001% by weight - solid).

**Figure 8** The structures of 5,6-dihydroxyindole (DHI) and 5,6-dihydroxyindole, 2-carboxylic acid (DHICA) – the monomeric precursors of eumelanin.

**Figure 9** Absorption coefficient at 380nm vs. concentration for the three synthetic eumelanin solutions in Fig. 7. Concentration errors were estimated to be $1 \times 10^{-4}$% by weight.

**Figure 10** Raw (A) and re-corrected (B) photoluminescence (PL) emission spectra (pumped at 380 nm) for three synthetic eumelanin solutions: 0.005% (dotted line), 0.0025% (dashed line) and 0.001% (solid line) by weight concentration, and solvent background (dot-dash line). (C) - Corrected PL emission peak intensity vs. concentration for the three synthetic eumelanin solutions.

**Figure 11** Re-corrected PL emission spectra for a 0.0025% by weight synthetic eumelanin solution: (A) plotted vs. wavelength, and (B) plotted vs. energy for five pump wavelengths: 360 nm (solid line) to 380 nm (inner dashed line) in 5 nm increments.

**Figure 12** Radiative relaxation quantum yield for synthetic eumelanin pumped at three wavelengths (350 nm, 380 nm and 410 nm). The solid line is a linear fit which is meant only as a guide to the eye.

**Figure 13** Time resolved PL emission spectrum of a 0.001% by weight synthetic eumelanin solution pumped at 395nm and monitored at 475nm. The response is clearly not single exponential, and the table shows the major lifetime components (with their relative amplitudes).

**Figure 14** The 3 redox forms of DHI – the indolequinone (IQ), semiquinone (SQ) and hydroquinone (HQ) tautomers.

**Figure 15** Conductivity (log) vs. relative humidity at room temperature for a solid pellet of synthetic eumelanin. Different relative humidity values were achieved using a range of saturated salt solutions, and the electrical measurements were made in a van der Pauw configuration.

**Figure 16** Photocurrent produced by a thin film of synthetic eumelanin produced by electropolymerisation. The photo-current was generated by illuminating the film with white light from a 150W Hg vapor lamp, and an electric field of ~65Vcm$^{-1}$ was used in a two electrode configuration.

**Figure 17** I-V response under AM1.5 illumination (150mWcm$^{-2}$) of a nano-crystalline titania DSSC with and without sensitization with a synthetic eumelanin. The sensitization was achieved using direct electropolymerisation of eumelanin onto the photoanode.

**Table 1**

| Compound | Extinction Coefficient at 380nm (g$^{-1}$cm$^{-1}$) | Extinction Coefficient at 518nm (g$^{-1}$cm$^{-1}$) |
|---|---|---|
| N3 | 17.9 | 42.0 |
| Synthetic eumelanin | 17.5 | 24.0 |

**Table 2**

| Molecule | TDDFT [46] (PBE/B3LYP) | TDDFT [46] (B3LYP) | ΔSCF (PBE) [36] | DFT (PBE) [36] | Huckel [40,36] |
|---|---|---|---|---|---|
| DHI | 4.53 | 4.30 | 3.61 | 3.48 | 3.40 |
| DHI (1,5) | 1.50 | 1.43 | 1.12 | 0.80 | 0.84 |
| DHI (5,6) | 1.82 | 1.79 | 2.02 | 1.07 | 1.30 |
| DHICA | - | - | 3.04 | 2.85 | - |
| DHICA (5) | - | - | 2.67 | 2.24 | - |
| DHICA (6) | - | - | 2.64 | 2.36 | - |
| DHICA (5,6) | - | - | 1.96 | 0.87 | - |
| DHICA (5,1) | - | - | 1.10 | 0.78 | - |
| DHICA (6,1) | - | - | 1.25 | 0.89 | - |

**Figure 1**

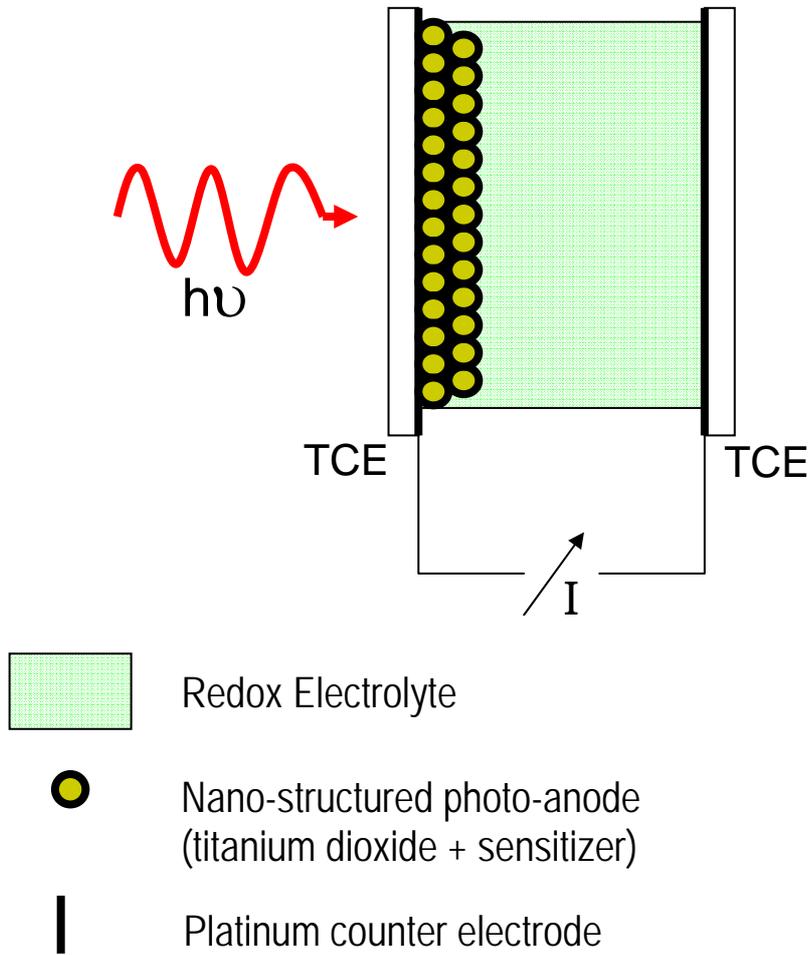

**Figure 2**

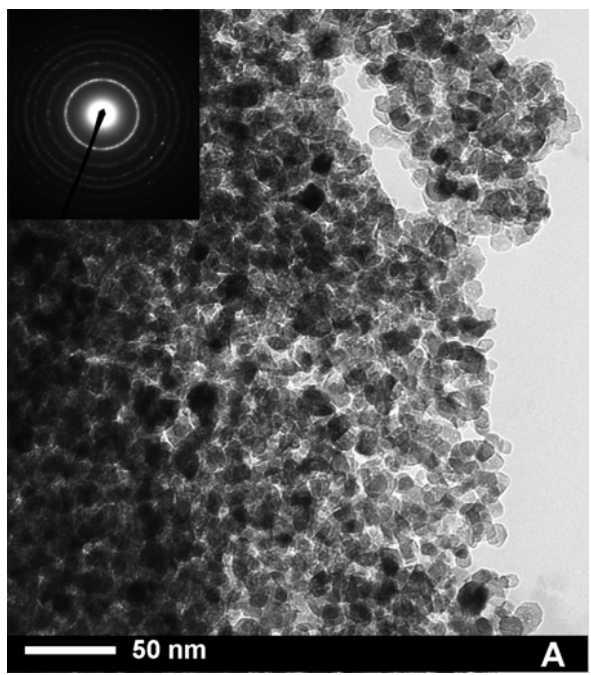

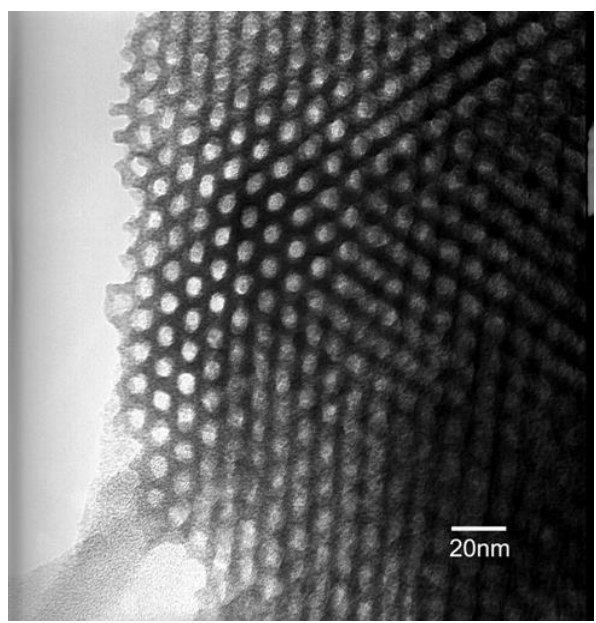

**Figure 3**

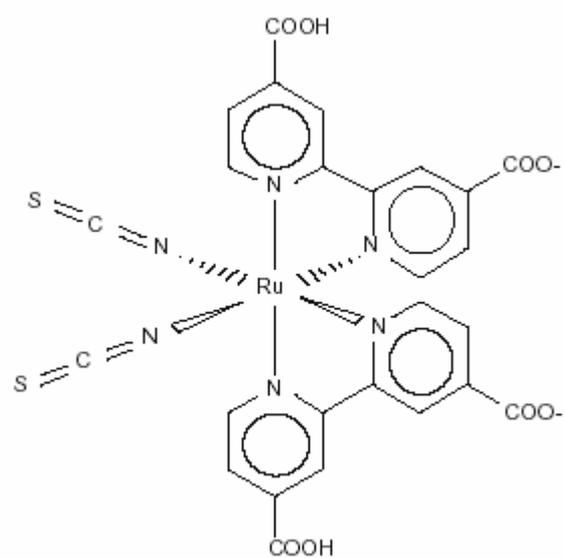

**Figure 4**

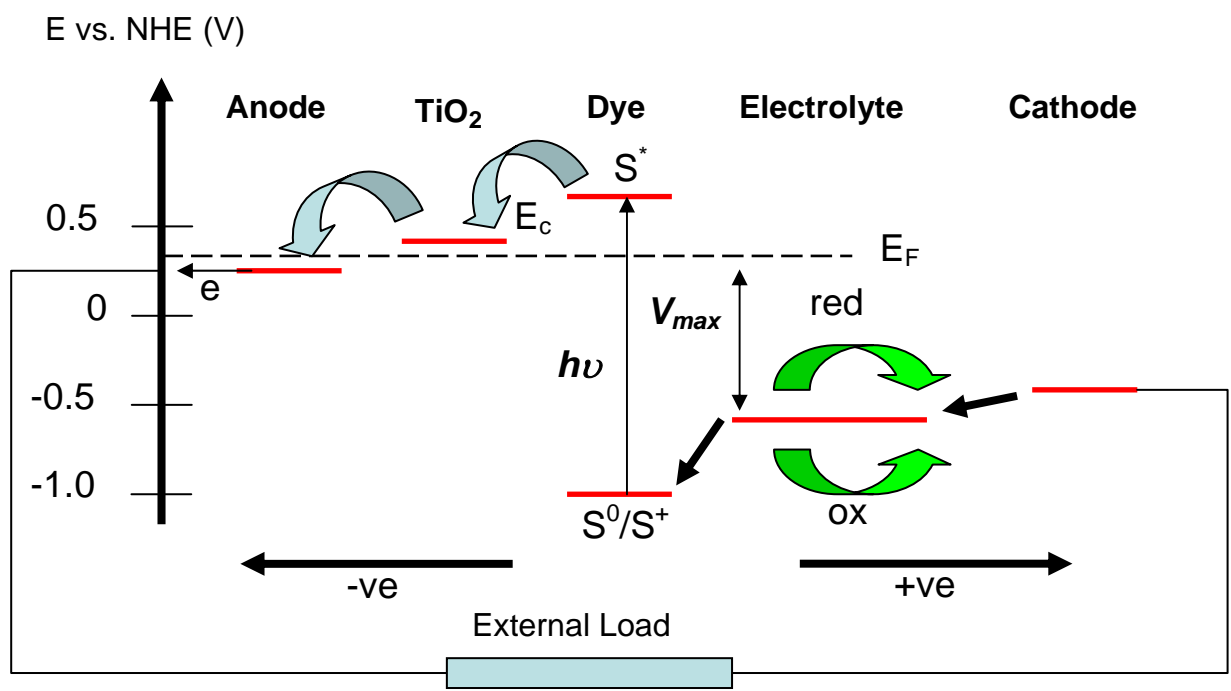

**Figure 5**

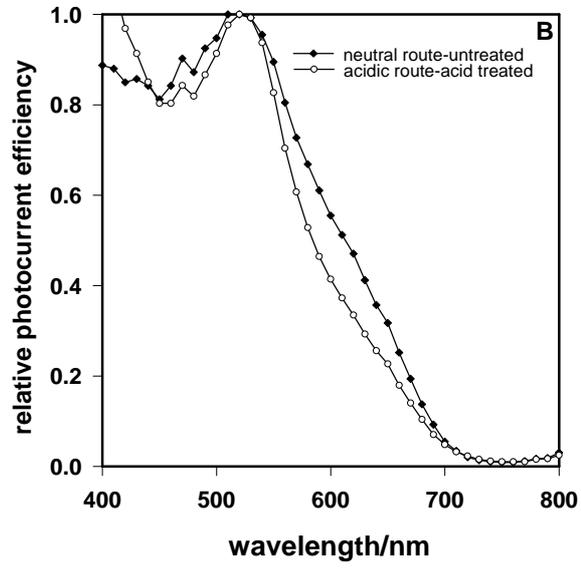

**Figure 6**

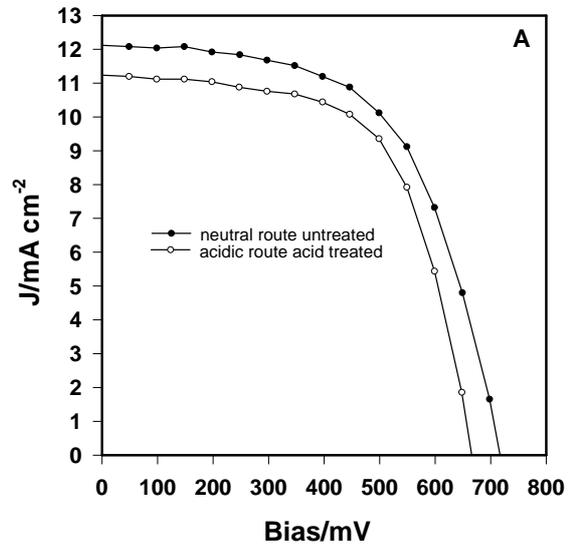

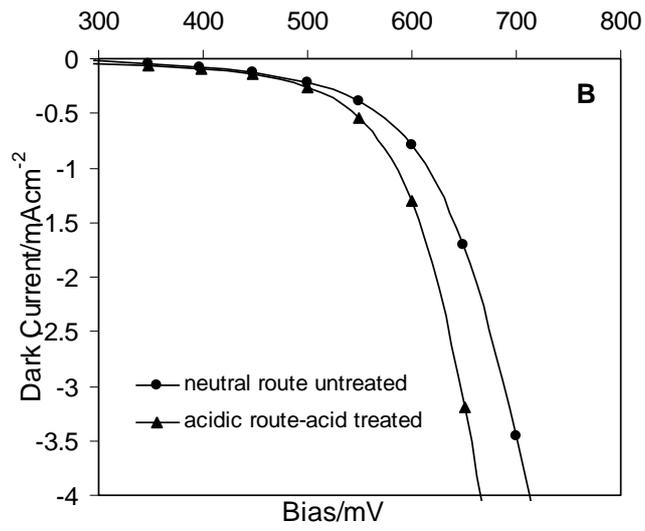

**Figure 7**

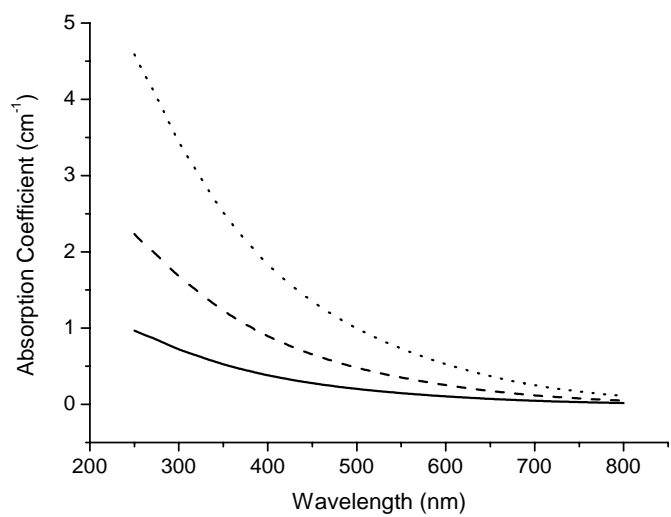

**Figure 8**

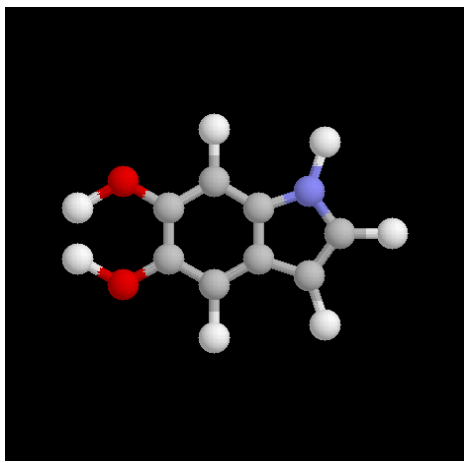

**DHI**

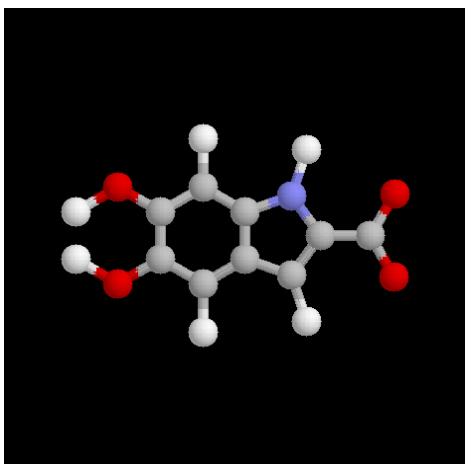

**DHICA**

**Figure 9**

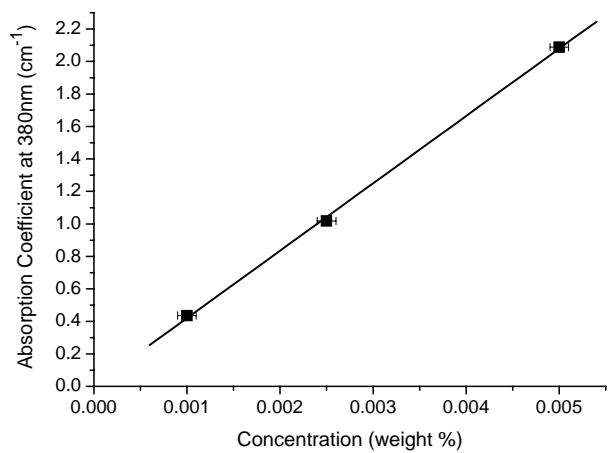

**Figure 10**

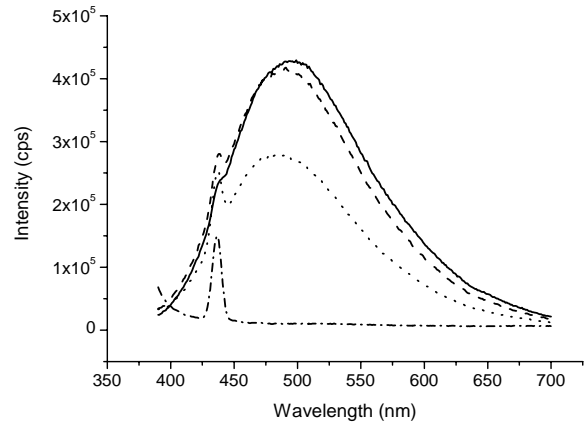

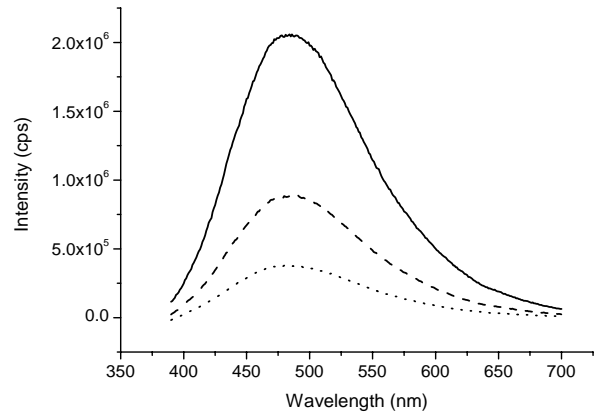

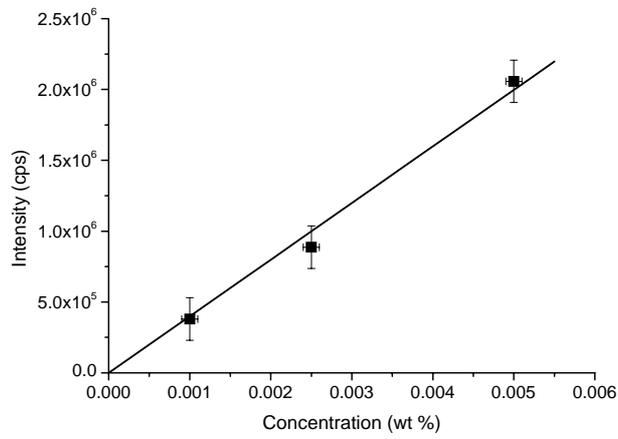

**Figure 11**

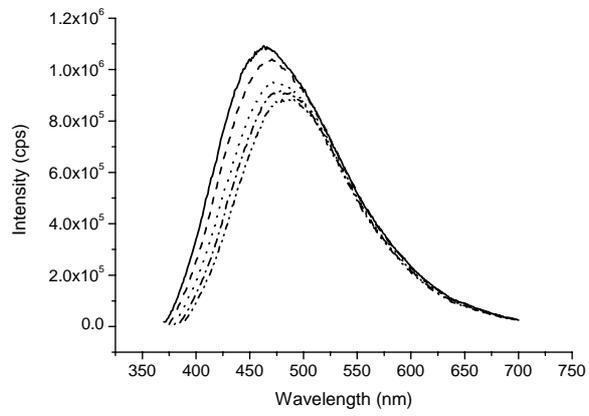

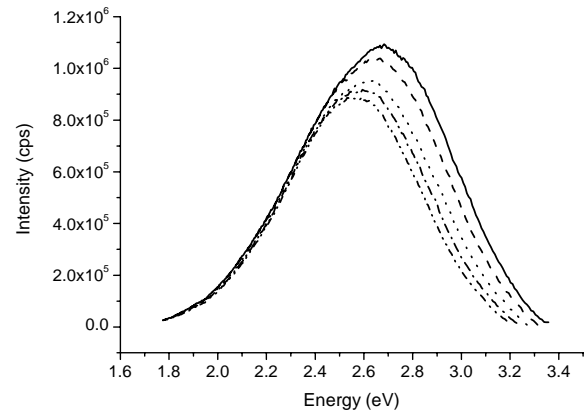

**Figure 12**

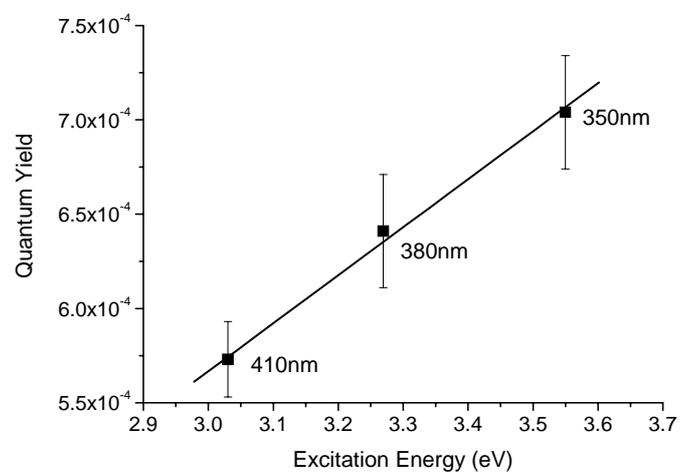

**Figure 13**

| Decay Time (ns) | Relative Amplitude (%) |
|---|---|
| 0.45 | 53 |
| 2.5 | 36 |
| 8.8 | 11 |

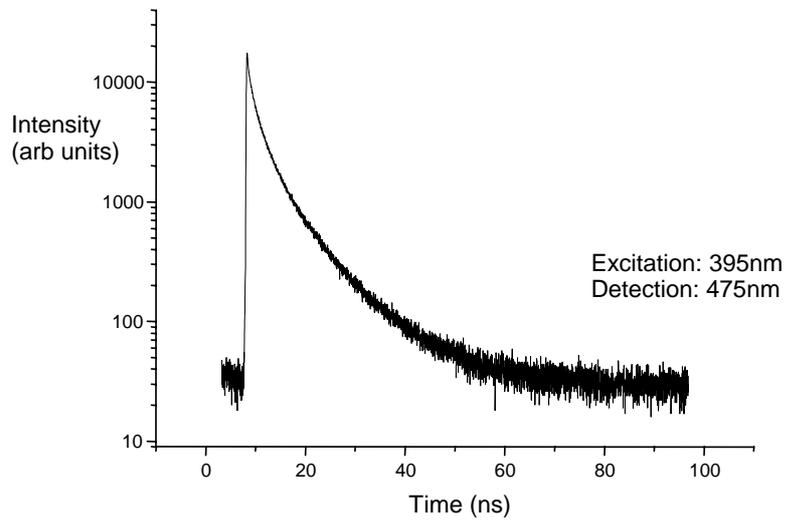

**Figure 14**

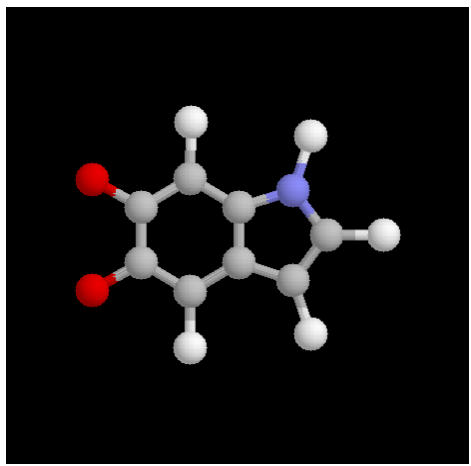

**IQ**

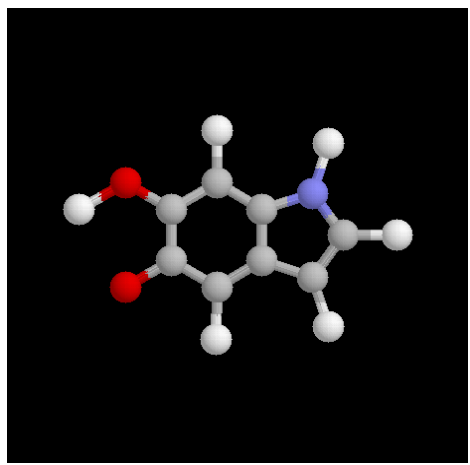

**SQ**

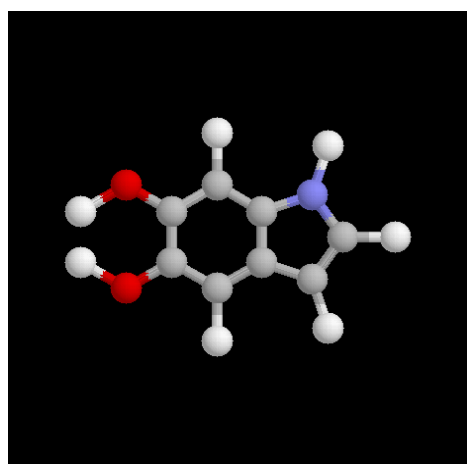

**HQ**

**Figure 15**

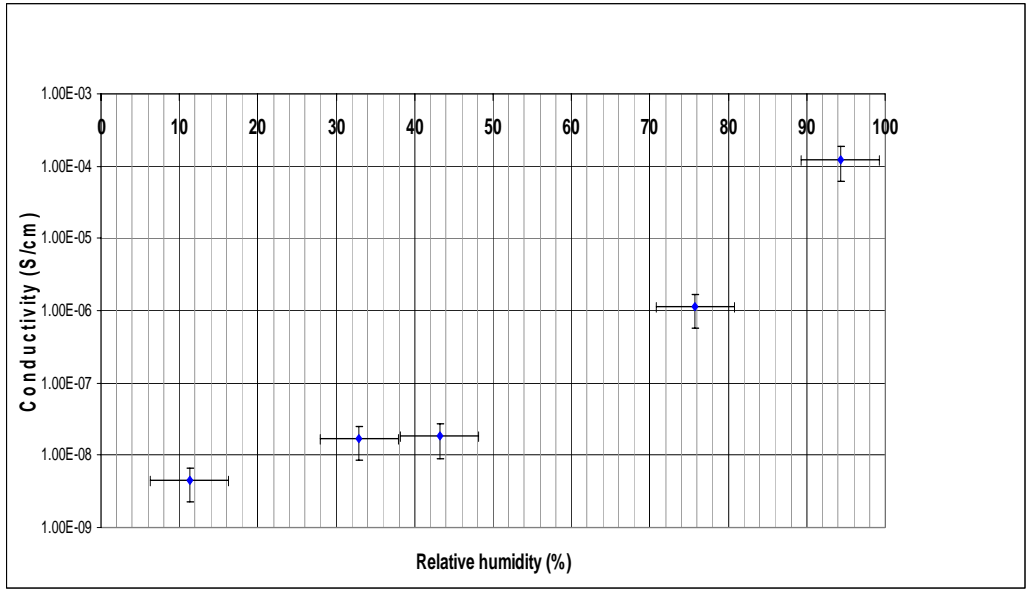

**Figure 16**

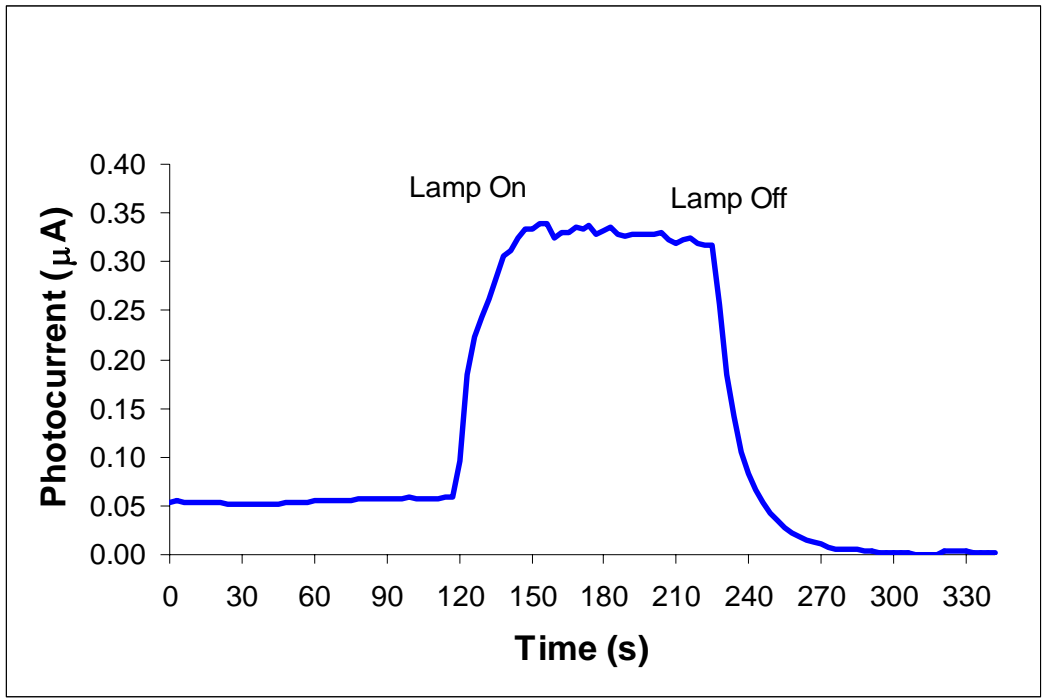

**Figure 17**

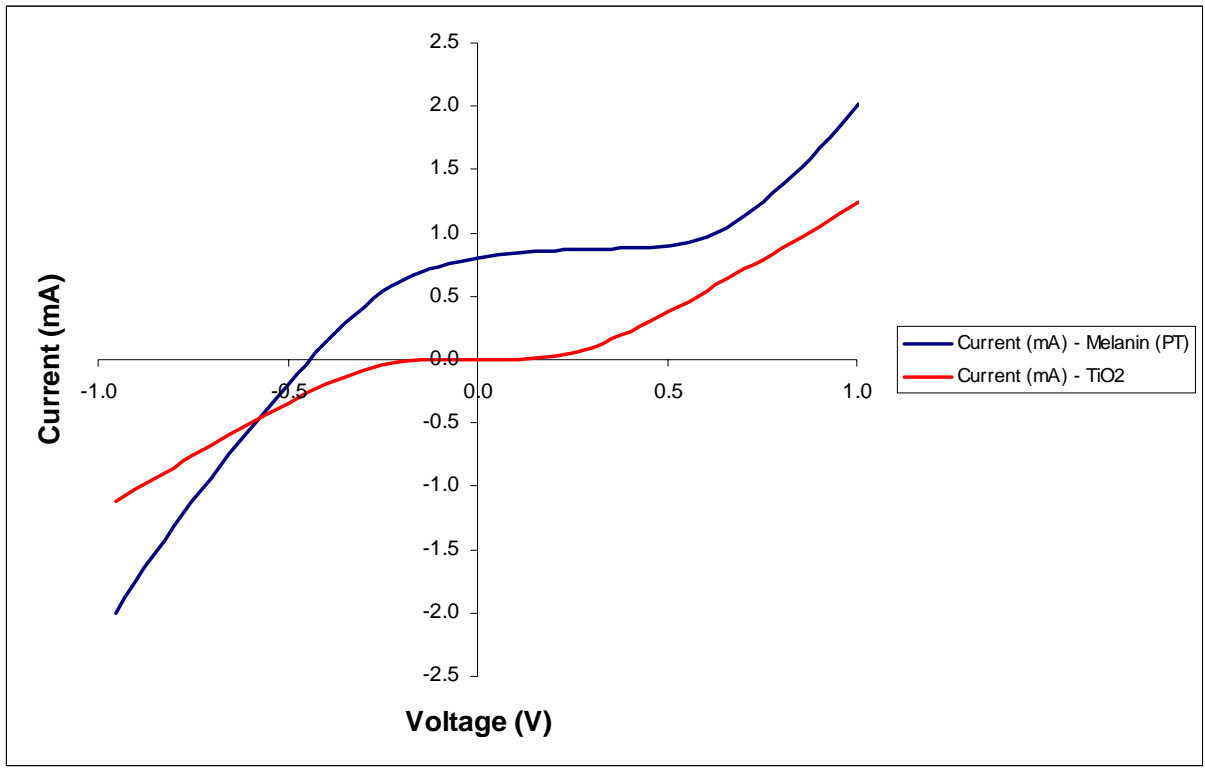